\documentclass[%
 reprint,
superscriptaddress,
 amsmath,amssymb,
 aps,
]{revtex4-1}

\usepackage{graphicx}
\usepackage{dcolumn}
\usepackage{bm}

\begin{document}

\preprint{APS/123-QED}

\title{Surface tension and contact with soft elastic solids}

\author{Robert W. Style}
\affiliation{%
Yale University, New Haven, CT 06520, USA 
}%

\author{Callen Hyland}%
\affiliation{%
Yale University, New Haven, CT 06520, USA 
}

\author{Rostislav Boltyanskiy}%
\affiliation{%
Yale University, New Haven, CT 06520, USA 
}

\author{John S. Wettlaufer}%
\affiliation{%
Yale University, New Haven, CT 06520, USA 
}
\affiliation{Mathematical Institute, University of Oxford, Oxford, OX1 3LB, UK}

\author{Eric R. Dufresne}%
\email[]{eric.dufresne@yale.edu}
\affiliation{%
Yale University, New Haven, CT 06520, USA 
}
%

%

\date{\today}

\begin{abstract}
Johnson-Kendall-Robert (JKR) theory is the basis of modern contact mechanics. It describes how two deformable objects adhere together, driven by adhesion energy and opposed by elasticity. However, it does not include solid surface tension, which also opposes adhesion by acting to flatten the surface of soft solids. We tested JKR theory to see if solid surface tension affects indentation behaviour. Using confocal microscopy, we characterised the indentation of glass particles into soft, silicone substrates. While JKR theory held for particles larger than a critical, elastocapillary lengthscale, it failed for smaller particles. Instead, adhesion of small particles mimicked the adsorption of particles at a fluid interface, with a size-independent contact angle between the undeformed surface and the particle given by a generalised version of Young's law. A simple theory quantitatively captures this behaviour, and explains how solid surface tension dominates elasticity for small-scale indentation of soft materials. 
\end{abstract}

\pacs{Valid PACS appear here}
\maketitle


Contacts between solid surfaces are found throughout nature and play important roles in almost every scientific field from physics \cite{john87,luan05,sure01} and biology \cite{mait12,arzt03} to astrophysics \cite{domi97,wett10} and meteorology \cite{reyn57,sher06}.
From an engineering perspective, an understanding of contacts is essential to control friction and adhesion \cite{mo09,land90,bran04,john71,mary06,chat10}.
The current understanding of soft contacts is based on the theory of Johnson-Kendall-Roberts (JKR)  -- which balances surface adhesive and bulk elastic energies \cite{john71}.
While JKR theory was originally developed to describe macroscopic contacts, it has been widely applied to microscopic systems \cite{butt05,erat10,noy97,luan05}.
Here, we show that JKR breaks down when the contact radius is smaller than a critical lengthscale determined by a balance of solid surface tension and elasticity. 
Instead,  small contacts are described by a generalisation of Young's law for liquid wetting on stiff solids  \cite{dege04}.  
Our results have important implications for any processes involving small contacts on soft materials including elastomers, gels, tissues and cells.
For example, JKR may not be appropriate for interpreting nanoindentation or atomic force microscopy data for soft materials.
Finally, solid surface tension may impact adhesion and friction of rough and hierarchical surfaces.

Adhesion and friction between  macroscopic solid surfaces are typically dominated by  the contact of  microscopic asperities \cite{bhus95,delr05,mo09}.
Intimate contact between the two solids liberates  free energy per unit area, $W$, called the adhesion energy.  
To increase  contact area,  adhering solids will deform.   
In the Johnson-Kendall-Roberts (JKR) \cite{john71} theory of contact  mechanics, the equilibrium  contact area is determined by balancing adhesion energy favouring contact against elastic energy opposing deformation.
JKR  is the basis of modern contact mechanics, and is applied across a wide variety of scientific and engineering disciplines.  
Here we show that JKR theory fails for small contacts on soft materials because it ignores solid surface tension.

Recent experiments have illustrated that solid surface tension, $\Upsilon_{sv}$, can play a dominant role in the behaviour of soft materials \cite{long96,jeri11,styl12,styl12c,styl13,chak13,mora13}.
Solid surface tension drives a rippling instability in soft, elongated structures \cite{mora10}.  
It smooths out sharp features,  limiting the resolution of lithography in gels and elastomers \cite{gord08,pers10,jago12}.
It also determines the wavelength of surface creases and ripples in a compressed gel \cite{mora11,chen12}.

These solid capillary effects become significant below a critical \emph{elastocapillary} lengthscale, $L$ \cite{mora10,jago12}.
The basic physics is highlighted by the following argument:
Consider a surface with a sinusoidal corrugation of wavelength $\lambda$.
Surface tension acts to flatten the surface, with  a  stress that scales like $\Upsilon_{sv}/\lambda^2$.
On the other hand, elastic forces will resist this deformation with a restoring stress that scales like $E/\lambda$, where $E$ is Young's modulus. 
When $\lambda \ll L = \Upsilon_{sv}/E$, solid surface tension overpowers elastic restoring forces and flattens the surface. 
Solid capillarity can be seen at the micron-scale for gels, at the nanometre-scale for elastomers, and is unimportant for harder materials such as glass \cite{styl12,styl12c,lee12,xu13}.
This scaling suggests that solid surface tension, ignored by JKR, may dominate small contacts with soft solids.

We tested the validity of JKR theory  by using confocal microscopy to measure the spontaneous indentation of hard, silica microspheres into soft, sticky, flat, silicone substrates.
We varied the particle size (3-30$\mu$m radii) and substrate stiffnesses ($E=3,\,85,\,150,\,500$kPa), and compared the resulting indentation profiles with theoretical predictions.

 \begin{figure}
\centering
  \includegraphics[width=9cm]{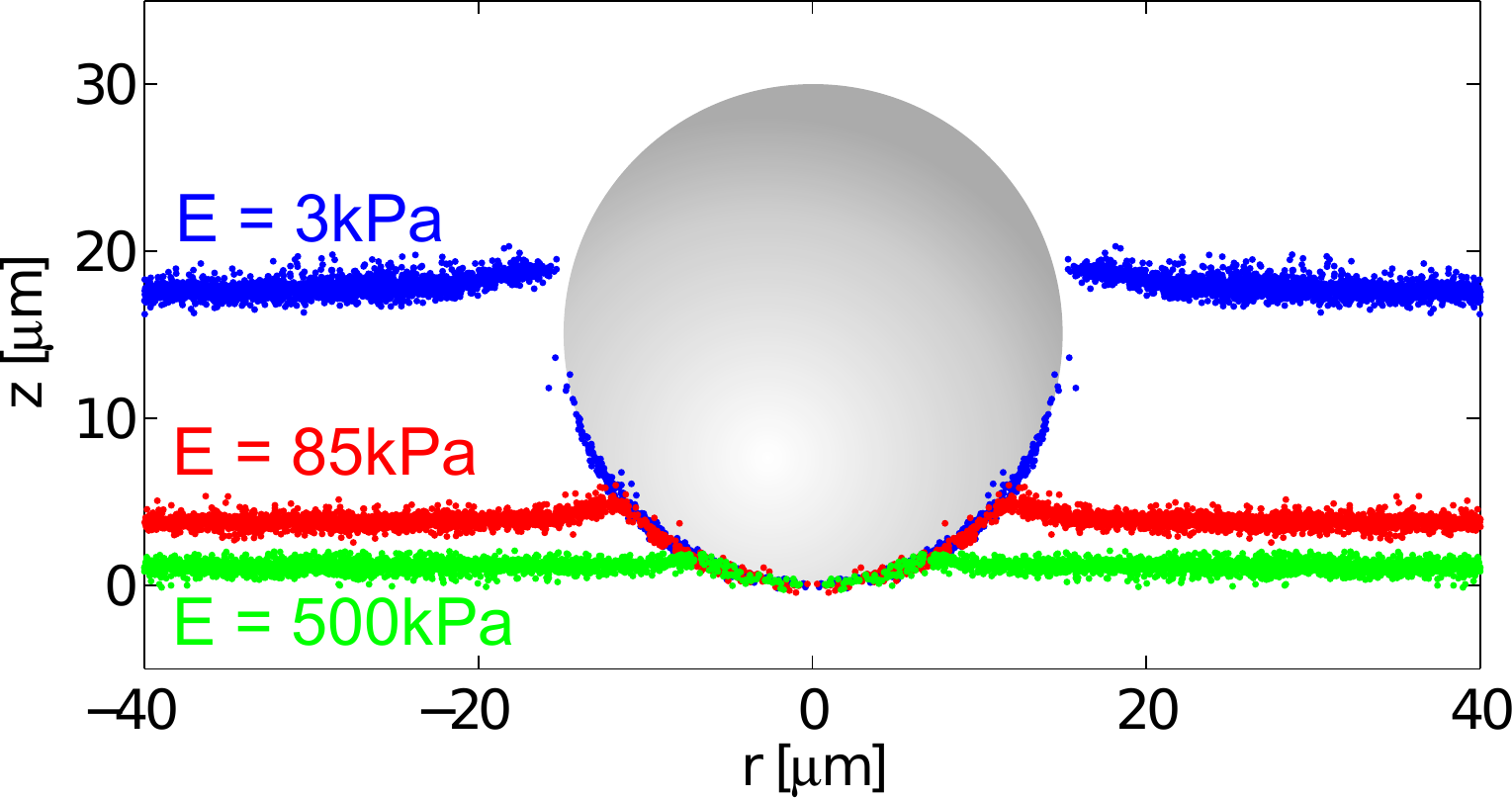}
  \caption{Surface profiles of silicone substrates of differing stiffness adhering to a glass microsphere of radius 15$\mu$m.  No external forces are applied to the particle except for its weight, which causes negligible indentation. Substrate stiffnesses are 3kPa (Blue), 85kPa (Red) and 500kPa (Green).}
  \label{fig:ex_profiles}
\end{figure}

Example surface profiles under 15$\mu$m spherical particles for three different stiffnesses are shown in Figure 1.
In each profile, the particle indents the substrate, with the indentation depth decreasing with substrate stiffness.
The measured indentation is entirely due to substrate-particle adhesion -- Hertz theory predicts that the  weight of the particles causes sub-nanometre indentations \cite{hert82}.
The substrate surface is also pulled up, adhering to the particle sides and creating a ridge at the contact line.
The clearest ridge appears on the intermediate 85kPa surface.
For the 3kPa substrate, the surface outside of the particle is surprisingly flat despite very large deformations under the bead, reminiscent of a particle adsorbed at a liquid surface \cite{dege04}.
Despite this `liquid-like' behaviour of the gel, it exhibits no plastic deformation at these strains  \cite{fren45,jago98,lin01},  as shown by rheology in the Supplemental Information.
Note that for the softest substrates, we could not  image the contact line completely, as the contact line is on the top half of the bead and  the objective is below the substrate.
Qualitatively similar profiles are seen for all 110 particles examined -- each shown in the Supplemental Information.

We extracted the indentation $d$ and contact radii $a$ of the particles from the surface profiles.
As shown in Figure 2, $d$, is the depth of the bottom of the particle relative to the undeformed substrate surface, and $a$ is the radius of the circular contact line.
Figure 2(a) shows $d$ vs particle radius, $R$, for different substrate stiffnesses.
Indentation increases with particle size and decreases with substrate stiffness.
Figure 2(b) shows the contact radii for the same experiments.
Contact radii similarly increase with particle size and decrease with  substrate stiffness.

\begin{figure}
\centering
  \includegraphics[width=9cm]{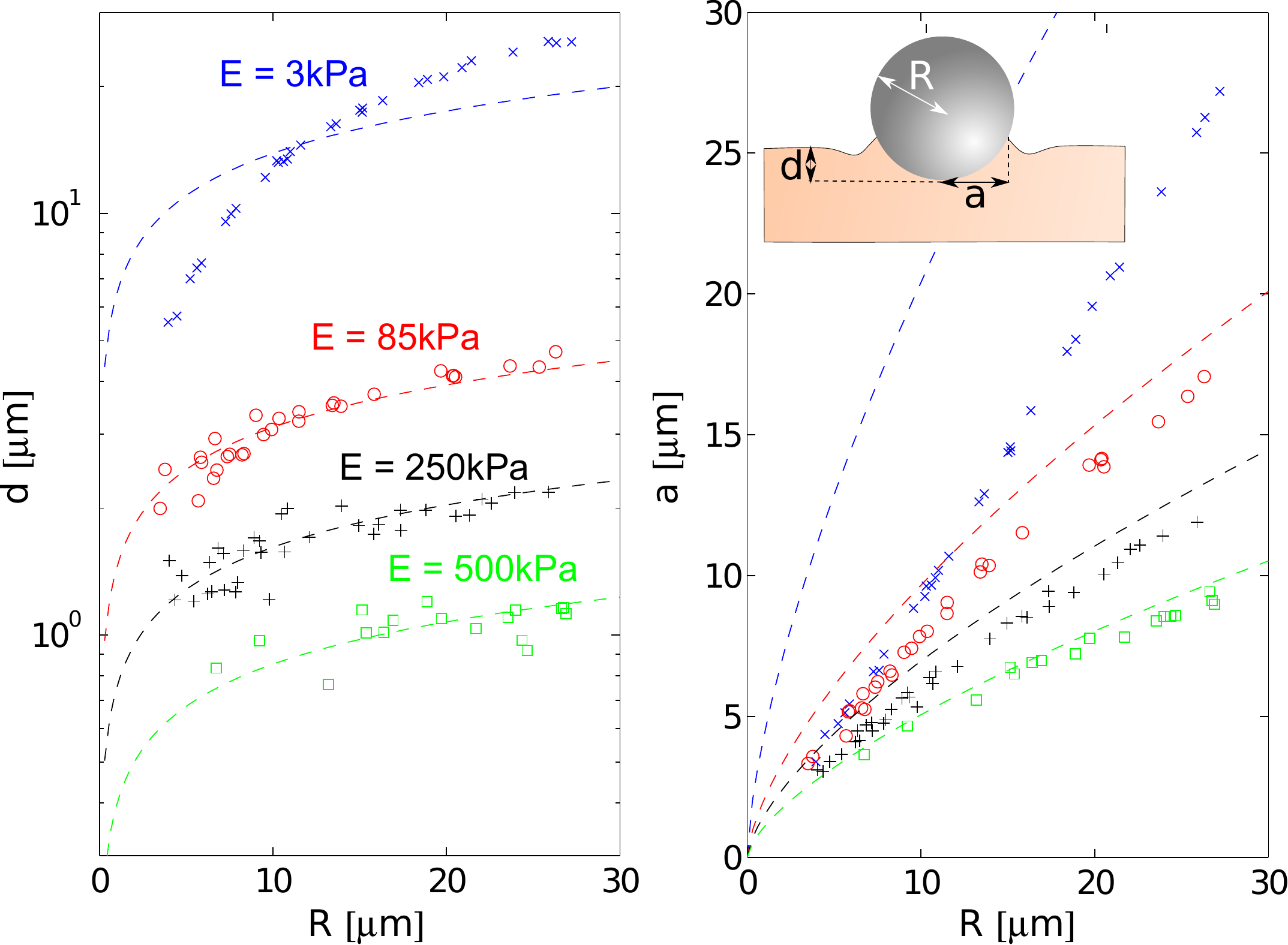}
  \caption{(a) Indentation depth vs particle radius and (b) contact radius vs particle radius for silica microspheres on silicone substrates of differing stiffnesses. Points show measured data for substrates with stiffness 3kPa (Blue), 85kPa (Red), 250kPa (Black) and 500kPa (Green). Dashed curves are the best fit predictions from JKR theory. Inset: Schematic showing the definitions of indentation, $d$, contact radius, $a$, and particle radius $R$.}  \label{fig:data}
\end{figure}

\begin{figure}
\centering
  \includegraphics[width=9cm]{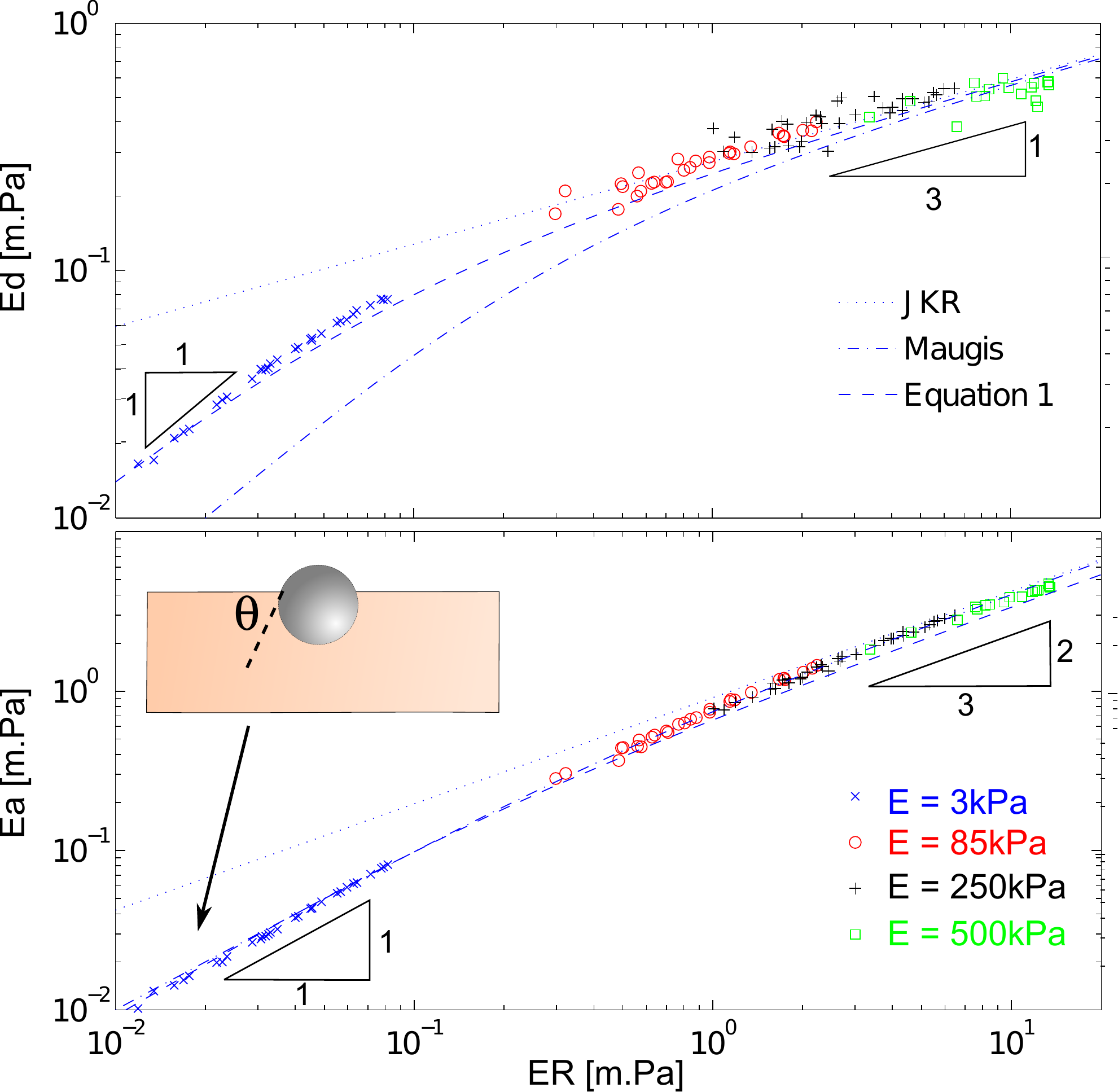}
  \caption{(a) Indentation $\times$ Young's modulus versus bead radius $\times$ Young's modulus. (b) Contact radius $\times$ Young's modulus versus bead radius $\times$ Young's modulus. Inset: schematic of particle behaviour for $ER\ll1$. For both quantities, the data collapses onto a smooth curve. The dotted line shows the best-fit JKR theory \cite{john71}. The dash-dotted curve shows Maugis's extended JKR predictions \cite{maug95}. The dashed curve is derived from equation \ref{eqn:energy}. Points show measured data for substrates with stiffness 3kPa (Blue), 85kPa (Red), 250kPa (Black) and 500kPa (Green).}
  \label{fig:collapse}
\end{figure}

Comparison of these results with JKR theory reveals significant discrepancies.
JKR theory applies for particles on elastic substrates when $d$ is much bigger than intermolecular distances \cite{tabo77,maug92}, and  clearly should be applicable to our data.
In the absence of an external load, it predicts an indentation $d=[\sqrt{3}\pi W(1-\nu^2)/2E]^{2/3}R^{1/3}$, and contact radius $a=[9\pi W(1-\nu^2)/2E]^{1/3}R^{2/3}$.
Here, $\nu$ is the substrate's Poisson ratio and the adhesion energy is defined as $W=\gamma_{sv} + \gamma_{pv} - \gamma_{sp}$, the change in interfacial energy on particle adhesion, per unit area.
Subscripts $s$, $p$ and $v$ indicate the soft substrate,  hard particle, and vapour respectively.
Silicone is nearly incompressible, so $\nu=1/2$, and the only unknown is the adhesion energy $W$ -- expected to be nearly the same for all the substrates.
We fit the indentation to the JKR prediction for each of the substrates, as shown in Figure 2(a).
For the three stiffer substrates, JKR theory agrees well with the indentation data, giving similar values of the adhesion energy:  $W$=72, 80 and 61mN/m for the 85, 250 and 500kPa substrates respectively.
For the 3kPa substrates, even the best fit value of $W$=24mN/m shows strong systematic deviation of the data from JKR theory.
Measured contact radii are compared to  JKR theory with the same values of $W$ in Figure 2(b).
Again, JKR agrees reasonably well with the contact radii for the three stiffest substrates,  despite  a small systematic over-estimate of the contact radius.
However, JKR theory  works for neither  indentation nor contact radius when $E=3$kPa.

The precise conditions where JKR theory starts to break down can be found by collapsing the indentation and contact-radius data onto a single curve.
In contact-mechanics theory, $a$ and $d$ can only depend on $E$ and $R$ and the interfacial tensions/energies.
We expect these interfacial quantities -- such as $W$ and $\Upsilon_{sv}$ -- to be nearly identical for all of our silicone substrates.
Then dimensional analysis shows that $Ed/W$ and $Ea/W$ can only vary with $ER/W$, or alternatively $Ea$ and $Ed$ can only depend on $ER$.
Figures 3(a,b) show the smooth collapse of the data when scaled by $E$.
The collapsed curves show that for large beads on stiffer substrates ($ER\gtrsim 3 \times 10^{-1}$m.Pa), the data matches with JKR theory with $W=71$mN/m, the average of the adhesion energies determined earlier for the three stiffer substrates.
For $ER\lesssim 3 \times 10^{-1}$m.Pa, there is a smooth deviation from the JKR predictions to a regime where $d$ and $a$ appear proportional to particle radius.

If JKR fails at small $ER$, what is the governing physics at this scale? The indentation profiles and scaling offer two important clues.  First,  in the small $ER$ regime, the indentation and contact radius scale with the particle radius: $d,a\sim R$.
This suggests that the undeformed surface of the substrate intersects the indenting sphere at a size-independent contact angle, $\theta$,  as shown in the inset of Figure 3b.  
Second, despite very large indentations, the substrates show nearly flat surfaces, as seen in Figure 1 and in the Supplemental Information.
These two features, which directly contradict the predictions of JKR, are instead identical to the surface-tension dominated behaviour of stiff particles adsorbed at a liquid surface \cite{dege04}.
Thus our results suggest a crossover from elastic  to  capillary contacts as $ER$ becomes smaller.

We need to rule out two other causes to confirm that the small $ER$ behaviour is due to solid-surface tension:
First, JKR theory assumes a parabolic, rather than spherical, particle shape.
We considered Maugis's extension of JKR theory for large deformations  \cite{maug95}.
Maugis's predictions are shown as dash-dotted curves in Figures 3(a,b), with $W=71$mN/m.
Even though this extension of JKR gives good agreement with the contact radii, there is a large deviation from the indentation data  for small $ER$.
Maugis's theory also predicts qualitatively different profiles to those measured in the experiments: Figure 4 shows the substrate profiles from Figure 1 along with JKR and Maugis preditions with $W=71$mN/m.
For the stiffer substrates, both theories work well.
However for the soft substrate, the measured profiles differ substantially from the theoretical ones.
Most obviously, Maugis' theory predicts a sharp ridge at the contact line which is missing in the data.
The second possible reason for failure of JKR/Maugis theory is that they only apply for linear-elastic response of the substrate.
For the soft, gel substrate, deformations are clearly not small.
However, we found the gel shows a linear stress-strain relationship until over 100\% shear strain (see Supplemental Information), so Maugis' large-deformation linear-elastic theory should give a reasonable  approximation of the substrate surface profile.
The large qualitative differences between experiments and theory (Figure 4) indicate that, despite possible non-linear effects, solid surface tension is a dominant effect in flattening the surface and controlling the particle indention in the small $ER$ regime.

\begin{figure}
\centering
\includegraphics[width=7cm]{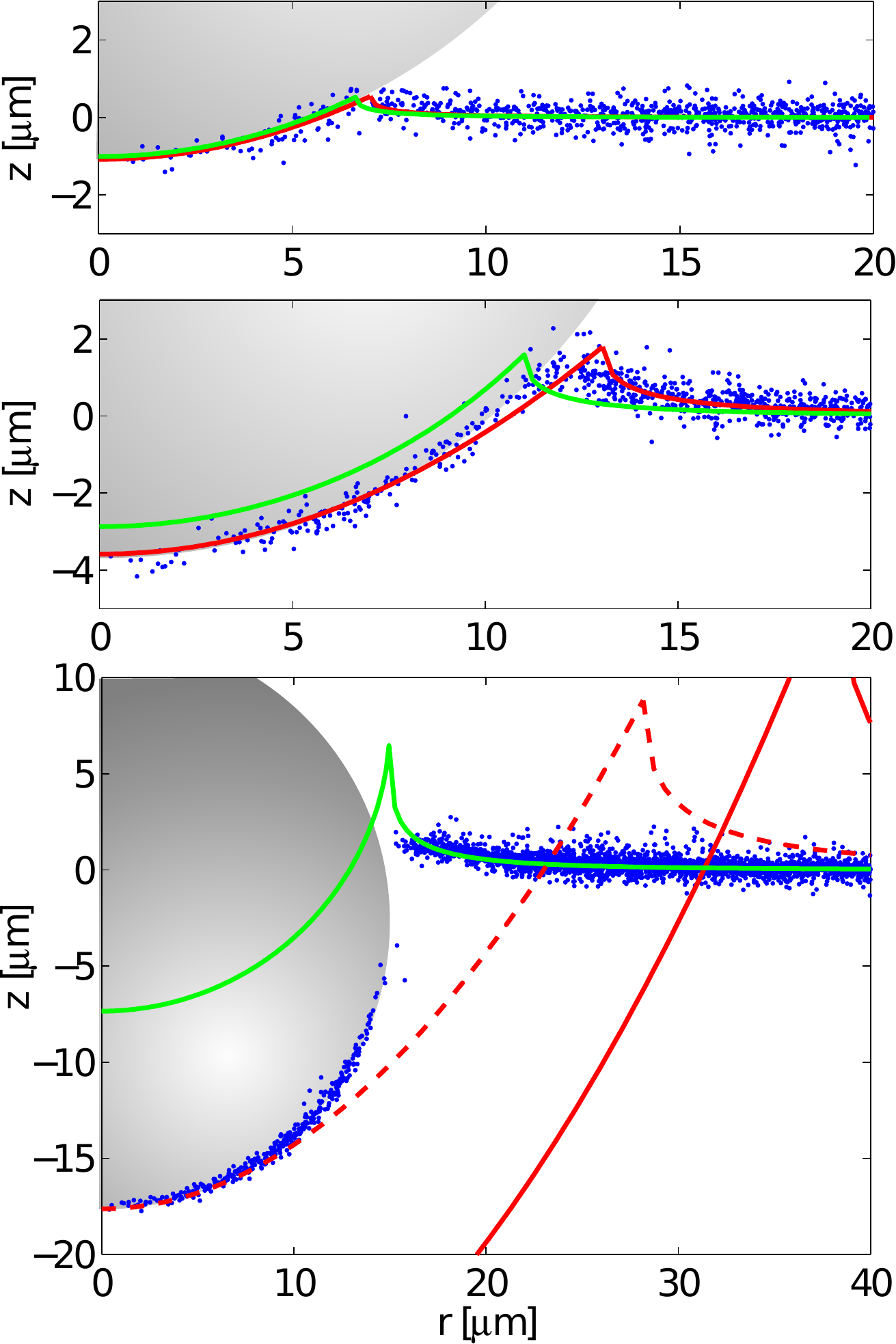}
  \caption{Surface profiles under 15$\mu$m radius beads compared with JKR theory (red curves) \cite{john71} and extended JKR theory (green curves)\cite{maug95} for $W=71$mN/m. From top to bottom: E=500kPa, E=85kPa and E=3kPa. Note the change in horizontal scale in the bottom figure. For the E=3kPa substrate, JKR theory is shown with $W=81$mN/m chosen to match the particle indentation (red-dashed).}
  \label{fig:profiles}
\end{figure}

The particle sizes and substrate stiffnesses where solid capillary effects arise can be calculated from a simple scaling argument.
The characteristic horizontal lengthscale in the system is min$(a,R)\sim $ min$((WR^2/E)^{1/3},R)$, where the scaling for $a$ comes from JKR theory.
Solid surface tension becomes important whenever this is smaller than $\Upsilon_{sv}/E$.
Thus we expect a capillary regime for $R\ll$max$(L,L\sqrt{\Upsilon_{sv}/W})$.
This is similar to the scaling found for theoretical predictions of nanoparticle adhesion on hard surfaces \cite{carr10}. 
For $\Upsilon_{sv}=30$mN/m \cite{styl12c} and $W=71$mN/m, the crossover will occur when $ER\sim \Upsilon_{sv}=30$mN/m, consistent with the transition behaviour seen in Figure 3.

A simple energy argument, modified from \cite{carr10} explains the transition from capillary to elastic regimes \cite{carr10}.
Breaking the particle adhesion process down into two stages, we first create a spherical indentation in the substrate, then we adhere the particle onto the indented area.
According to Hertz theory \cite{hert82}, the elastic energy required to make a spherical indentation of depth $d$ and radius of curvature $R$ is $U_{el} \sim ER^{1/2}d^{5/2}/(1-\nu^2)$.
Indenting the substrate also stretches the surface, creating new surface area. 
This introduces another energy penalty, ignored by JKR, equal to the surface tension times the additional surface area.
Approximating the indentation as a spherical cap in a flat plane gives $U_{\Upsilon}=\pi\Upsilon_{sv} d^2$.
Note, the surface tension, (also known as surface stress), $\Upsilon_{sv}$, is not always equivalent to the interfacial energy $\gamma_{sv}$ \cite{shut50,hui13}.
For solids, $\gamma_{sv}$ gives the work needed to create additional surface area by cleaving, while
$\Upsilon_{sv}$ gives the work needed to create additional surface area by stretching.
For fluids, $\gamma_{sv}=\Upsilon_{sv}$.
The final contribution to the total energy comes from  adhesion of the particle to the stretched substrate.
This is the term that drives indentation and is equal to $W$ times the adhered area, $U_{ad}=-2\pi WRd$ (again we approximate the indentation as a spherical cap in a flat plane).
Thus, the total energy change upon indentation is
\begin{equation}
U=cER^{1/2}d^{5/2}/(1-\nu^2)+\pi\Upsilon_{sv} d^2-2\pi WRd,
\end{equation}
where $c$ is a constant to be determined.

Minimising $U$ with respect to $d$, we obtain
\begin{equation}
\frac{5cER^{1/2}d^{3/2}}{2(1-\nu^2)}+2\pi \Upsilon_{sv} d-2\pi WR=0.
\label{eqn:energy}
\end{equation}
For large $R$, adhesion is balanced by elasticity, and we recover JKR theory with $c=8/(5\sqrt{3})$.
This is shown schematically in Figure 5(a).
For small $R$, the elastic response of the substrate falls out and adhesion is balanced by substrate surface tension, with Equation \ref{eqn:energy} reducing to  $d=WR/\Upsilon_{sv}$.
Equivalently,
\begin{equation}
\Upsilon_{sv}\cos\theta=W-\Upsilon_{sv}=\gamma_{pv}-(\gamma_{sp}+[\Upsilon_{sv}-\gamma_{sv}]),
\label{eqn:yl}
\end{equation}
where $\theta$ is the angle of the sphere's surface relative to the underformed substrate at the contact line, analogous to the contact angle in wetting.
This result reduces to Young's law when the substrate has a fluid-like surface tension $\Upsilon_{sv}=\gamma_{sv}$. 
More generally, we find that the adhesion of small spheres to soft surfaces is formally identical to the adsorption hard particles to a fluid interface (Figure 5).

\begin{figure}
\centering
\includegraphics[width=7cm]{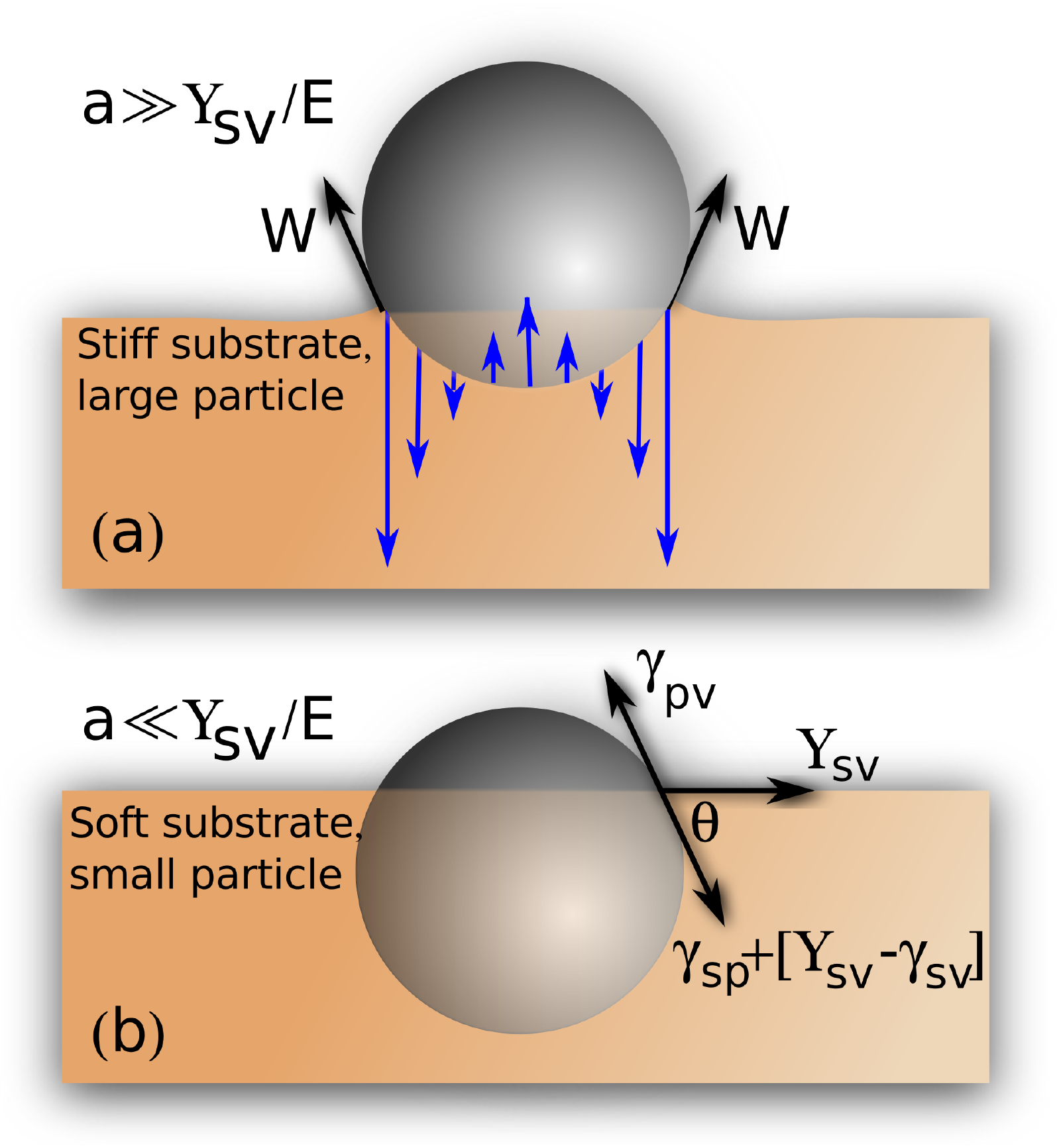}
  \caption{Hard and soft contacts. (a) When the contact radius is much larger than $\Upsilon_{sv}/E$, the contact is described by JKR theory which balances adhesion  and elastic stresses  (blue arrows).  The surface tension contribution is negligible. (b) When the contact radius is much smaller than $\Upsilon_{sv}/E$, surface tension governs contact mechanics and adhesion mimics adsorption onto a fluid interface.  The undeformed solid surface hits the particle at a fixed contact angle given by the generalised Young's law in Equation \ref{eqn:yl}. Elastic forces are negligible.}
  \label{fig:schem}
\end{figure}

Our indentation data is well described by the simple energy scaling embodied in Equation \ref{eqn:energy}.
We use the measured value of $W=71$mN/m and fit $\Upsilon_{sv}=45$mN/m.
The results are in excellent agreement with the data in Figure 3(a).
The contact radius is estimated from the spherical-cap relationship $a=\sqrt{2Rh-h^2}$.
 This systematically underestimates the contact radius in the elastic regime, since we ignore the adhesive ridge.
 Despite this, it nicely captures the transition from capillary to elastic regimes in Figure 3(b).
The extracted value of $\Upsilon_{sv}$ is a similar magnitude to previously measured values of $\Upsilon_{sv}=31\pm 5$mN/m \cite{styl12c}.
A caveat is that our scaling uses a linear elastic approximation for the elastic energy.
This does not affect the accuracy of theoretical predictions in the JKR, or capillary-dominated limits.
However, it may lead to inaccuracies in predictions of indentations and contact radii in the transition region where strains are large, and elastic energy is important.

These results demonstrate a new technique for measuring surface stresses in soft solids by measuring particle indentations.
Indentation with a large particle in the JKR regime gives the adhesion energy, $W$.
Indentation with a small particle in the capillary regime gives the surface tension $\Upsilon_{sv}$ from the relationship $\Upsilon_{sv}=WR/d$.
There are currently only a few, recently-developed techniques for measurement of solid surface tensions in soft surfaces.
These either require imaging of surface profiles close to a droplet contact line \cite{styl12c}, detailed elastic models \cite{jago12}, or bending of elastic membranes on wetting \cite{nade13}.
This indentation offers a new, simpler method for dry measurement of surface tensions.

The mechanical characterisation of soft substrates by nanoindentation and Atomic Force Microscopy (AFM) is likely very sensitive to  surface tension. 
The curvature of an AFM tip can be just a few nanometers.
There solid surface tension will certainly affect indentations of gels and elastomers where the elastocapillary lengths are on the scale of tens of microns or tens of nanometers, respectively. 
On the other hand, the effect of solid surface tension can be reduced by submerging the substrate in a fluid that lowers its surface tension,  reducing the critical lengthscale $L$.
Note also that in the limit of very small indenters, effects such as line tension or the breakdown of continuum theory may also become important \cite{schi07}.

In conclusion, JKR contact theory breaks down when contact radii are smaller than a critical elastocapillary length.
In this regime, the adhesion of stiff particles to soft solids mimics the adsorption of particles at a fluid interface.
Consequently,  solid surface tension must be accounted for in the interpretation of AFM and nano-indentation data on a variety of soft surfaces.
Our theoretical predictions match well with measured indentation depths for force-free indenters, and suggest a new, straightforward method of measuring solid surface tensions via indentation.
Our results should be extended to externally-forced particles; whenever min$(a,R)\sim\Upsilon_{sv}/E$ , JKR or Hertz theories will not hold due to significant solid surface tension effects.
Future work should consider contact between two soft solids.
In that case, for small particle contacts, we anticipate similar behaviour to the wetting of a liquid droplet on an an immiscible fluid substrate. 
Our results elucidate a fundamental difference in the contact mechanics of soft and hard materials, summarised in Figure 5, and  may lead to novel approaches to engineering adhesion and friction.

RWS is funded by the Yale University Bateman Interdepartmental Postdoctoral Fellowship. JSW thanks the Swedish Research Council for support. Support for instrumentation was provided by NSF (DBI-0619674).  We thank Anand Jagota and Chung-Yuen Hui for helpful feedback.

{\footnotesize \section*{Methods}
Soft adhesive substrates were created by spin-coating  silicone gel and elastomer onto glass slides. We prepared very soft substrates with thickness $h=150\mu$m from a silicone gel (CY 52-276 A/B, Dow Corning) with Young's modulus $E=$3kPa.
Rheometry shows that this gel behaves elastically to over 100\% strain with a constant Young's modulus (see Supplemental Information).
We made stiffer substrates with $h=70-72\mu$m and with $E=85$, 250 and 500kPa from a silicone elastomer (Sylgard 184, Dow Corning).
This elastomer is also highly-elastic, with rheology documented in the literature (e.g. \cite{chen13}).
Young's moduli for the elastomer substrates were interpolated from previously reported data \cite{cesa07,fuar08,merk07}.
To image substrate topography, we covalently attached fluorescent nanobeads (40nm carboxylated Yellow-Green Fluospheres, Invitrogen) to the substrates \cite{xu10}.
These covered a total area fraction of $<10^{-3}$.
To facilitate nanobead attachment, the substrates were exposed to (3-Aminopropyl)triethoxysilane vapour for 2 minutes.

We quantified the deformation of substrates after adhesion to microscopic glass spheres. We distributed $3-30\mu$m radii  glass spheres (Polysciences) well-apart on the substrate, and recorded the positions of the underlying fluorescent nanobeads by confocal microscopy with a 60x, NA 1.2 objective \cite{jeri11,styl12c}.
Nanobead positions were then extracted by image analysis, and azimuthally-collapsed to give radial surface profiles under each silica particle, as described in \cite{styl12c}, and detailed further in the Supplemental Information.
From the radial profiles, we determined silica particle radii by fitting a circle through the adhered section of the substrate surface.
These values agreed with particle radii measured by brightfield imaging.
In total we measured 110 particles on the four different substrate stiffnesses, and ensured that substrates were much thicker than the contact radii between the particles and substrate.
}


\begin{thebibliography}{53}%
\makeatletter
\providecommand \@ifxundefined [1]{%
 \@ifx{#1\undefined}
}%
\providecommand \@ifnum [1]{%
 \ifnum #1\expandafter \@firstoftwo
 \else \expandafter \@secondoftwo
 \fi
}%
\providecommand \@ifx [1]{%
 \ifx #1\expandafter \@firstoftwo
 \else \expandafter \@secondoftwo
 \fi
}%
\providecommand \natexlab [1]{#1}%
\providecommand \enquote  [1]{``#1''}%
\providecommand \bibnamefont  [1]{#1}%
\providecommand \bibfnamefont [1]{#1}%
\providecommand \citenamefont [1]{#1}%
\providecommand \href@noop [0]{\@secondoftwo}%
\providecommand \href [0]{\begingroup \@sanitize@url \@href}%
\providecommand \@href[1]{\@@startlink{#1}\@@href}%
\providecommand \@@href[1]{\endgroup#1\@@endlink}%
\providecommand \@sanitize@url [0]{\catcode `\\12\catcode `\$12\catcode
  `\&12\catcode `\#12\catcode `\^12\catcode `\_12\catcode `\%12\relax}%
\providecommand \@@startlink[1]{}%
\providecommand \@@endlink[0]{}%
\providecommand \url  [0]{\begingroup\@sanitize@url \@url }%
\providecommand \@url [1]{\endgroup\@href {#1}{\urlprefix }}%
\providecommand \urlprefix  [0]{URL }%
\providecommand \Eprint [0]{\href }%
\providecommand \doibase [0]{http://dx.doi.org/}%
\providecommand \selectlanguage [0]{\@gobble}%
\providecommand \bibinfo  [0]{\@secondoftwo}%
\providecommand \bibfield  [0]{\@secondoftwo}%
\providecommand \translation [1]{[#1]}%
\providecommand \BibitemOpen [0]{}%
\providecommand \bibitemStop [0]{}%
\providecommand \bibitemNoStop [0]{.\EOS\space}%
\providecommand \EOS [0]{\spacefactor3000\relax}%
\providecommand \BibitemShut  [1]{\csname bibitem#1\endcsname}%
\let\auto@bib@innerbib\@empty
\bibitem [{\citenamefont {Johnson}\ and\ \citenamefont
  {Johnson}(1987)}]{john87}%
  \BibitemOpen
  \bibfield  {author} {\bibinfo {author} {\bibfnamefont {K.~L.}\ \bibnamefont
  {Johnson}}\ and\ \bibinfo {author} {\bibfnamefont {K.~K.~L.}\ \bibnamefont
  {Johnson}},\ }\href@noop {} {\emph {\bibinfo {title} {Contact mechanics}}}\
  (\bibinfo  {publisher} {Cambridge university press},\ \bibinfo {year}
  {1987})\BibitemShut {NoStop}%
\bibitem [{\citenamefont {Luan}\ and\ \citenamefont {Robbins}(2005)}]{luan05}%
  \BibitemOpen
  \bibfield  {author} {\bibinfo {author} {\bibfnamefont {B.}~\bibnamefont
  {Luan}}\ and\ \bibinfo {author} {\bibfnamefont {M.~O.}\ \bibnamefont
  {Robbins}},\ }\href@noop {} {\bibfield  {journal} {\bibinfo  {journal}
  {Nature}\ }\textbf {\bibinfo {volume} {435}},\ \bibinfo {pages} {929}
  (\bibinfo {year} {2005})}\BibitemShut {NoStop}%
\bibitem [{\citenamefont {Suresh}(2001)}]{sure01}%
  \BibitemOpen
  \bibfield  {author} {\bibinfo {author} {\bibfnamefont {S.}~\bibnamefont
  {Suresh}},\ }\href@noop {} {\bibfield  {journal} {\bibinfo  {journal}
  {Science}\ }\textbf {\bibinfo {volume} {292}},\ \bibinfo {pages} {2447}
  (\bibinfo {year} {2001})}\BibitemShut {NoStop}%
\bibitem [{\citenamefont {Ma"tre}\ \emph {et~al.}(2012)\citenamefont {Ma"tre},
  \citenamefont {Berthoumieux}, \citenamefont {Krens}, \citenamefont
  {Salbreux}, \citenamefont {JŸlicher}, \citenamefont {Paluch},\ and\
  \citenamefont {Heisenberg}}]{mait12}%
  \BibitemOpen
  \bibfield  {author} {\bibinfo {author} {\bibfnamefont {J.-L.}\ \bibnamefont
  {Ma"tre}}, \bibinfo {author} {\bibfnamefont {H.}~\bibnamefont
  {Berthoumieux}}, \bibinfo {author} {\bibfnamefont {S.~F.~G.}\ \bibnamefont
  {Krens}}, \bibinfo {author} {\bibfnamefont {G.}~\bibnamefont {Salbreux}},
  \bibinfo {author} {\bibfnamefont {F.}~\bibnamefont {JŸlicher}}, \bibinfo
  {author} {\bibfnamefont {E.}~\bibnamefont {Paluch}}, \ and\ \bibinfo {author}
  {\bibfnamefont {C.-P.}\ \bibnamefont {Heisenberg}},\ }\href@noop {}
  {\bibfield  {journal} {\bibinfo  {journal} {Science}\ }\textbf {\bibinfo
  {volume} {338}},\ \bibinfo {pages} {253} (\bibinfo {year}
  {2012})}\BibitemShut {NoStop}%
\bibitem [{\citenamefont {Arzt}\ \emph {et~al.}(2003)\citenamefont {Arzt},
  \citenamefont {Gorb},\ and\ \citenamefont {Spolenak}}]{arzt03}%
  \BibitemOpen
  \bibfield  {author} {\bibinfo {author} {\bibfnamefont {E.}~\bibnamefont
  {Arzt}}, \bibinfo {author} {\bibfnamefont {S.}~\bibnamefont {Gorb}}, \ and\
  \bibinfo {author} {\bibfnamefont {R.}~\bibnamefont {Spolenak}},\ }\href@noop
  {} {\bibfield  {journal} {\bibinfo  {journal} {Proc. Nat. Acad. Sci.}\
  }\textbf {\bibinfo {volume} {100}},\ \bibinfo {pages} {10603} (\bibinfo
  {year} {2003})}\BibitemShut {NoStop}%
\bibitem [{\citenamefont {Dominik}\ and\ \citenamefont
  {Tielens}(1997)}]{domi97}%
  \BibitemOpen
  \bibfield  {author} {\bibinfo {author} {\bibfnamefont {C.}~\bibnamefont
  {Dominik}}\ and\ \bibinfo {author} {\bibfnamefont {A.}~\bibnamefont
  {Tielens}},\ }\href@noop {} {\bibfield  {journal} {\bibinfo  {journal}
  {Astrophys. J.}\ }\textbf {\bibinfo {volume} {480}},\ \bibinfo {pages} {647}
  (\bibinfo {year} {1997})}\BibitemShut {NoStop}%
\bibitem [{\citenamefont {Wettlaufer}(2010)}]{wett10}%
  \BibitemOpen
  \bibfield  {author} {\bibinfo {author} {\bibfnamefont {J.}~\bibnamefont
  {Wettlaufer}},\ }\href@noop {} {\bibfield  {journal} {\bibinfo  {journal}
  {Astrophys. J.}\ }\textbf {\bibinfo {volume} {719}},\ \bibinfo {pages} {540}
  (\bibinfo {year} {2010})}\BibitemShut {NoStop}%
\bibitem [{\citenamefont {Reynolds}\ \emph {et~al.}(1957)\citenamefont
  {Reynolds}, \citenamefont {Brook},\ and\ \citenamefont {Gourley}}]{reyn57}%
  \BibitemOpen
  \bibfield  {author} {\bibinfo {author} {\bibfnamefont {S.}~\bibnamefont
  {Reynolds}}, \bibinfo {author} {\bibfnamefont {M.}~\bibnamefont {Brook}}, \
  and\ \bibinfo {author} {\bibfnamefont {M.~F.}\ \bibnamefont {Gourley}},\
  }\href@noop {} {\bibfield  {journal} {\bibinfo  {journal} {J. Meteorol.}\
  }\textbf {\bibinfo {volume} {14}},\ \bibinfo {pages} {426} (\bibinfo {year}
  {1957})}\BibitemShut {NoStop}%
\bibitem [{\citenamefont {Sherwood}\ \emph {et~al.}(2006)\citenamefont
  {Sherwood}, \citenamefont {Phillips},\ and\ \citenamefont
  {Wettlaufer}}]{sher06}%
  \BibitemOpen
  \bibfield  {author} {\bibinfo {author} {\bibfnamefont {S.~C.}\ \bibnamefont
  {Sherwood}}, \bibinfo {author} {\bibfnamefont {V.~T.}\ \bibnamefont
  {Phillips}}, \ and\ \bibinfo {author} {\bibfnamefont {J.}~\bibnamefont
  {Wettlaufer}},\ }\href@noop {} {\bibfield  {journal} {\bibinfo  {journal}
  {Geophys. Res. Lett.}\ }\textbf {\bibinfo {volume} {33}},\ \bibinfo {pages}
  {L05804} (\bibinfo {year} {2006})}\BibitemShut {NoStop}%
\bibitem [{\citenamefont {Mo}\ \emph {et~al.}(2009)\citenamefont {Mo},
  \citenamefont {Turner},\ and\ \citenamefont {Szlufarska}}]{mo09}%
  \BibitemOpen
  \bibfield  {author} {\bibinfo {author} {\bibfnamefont {Y.}~\bibnamefont
  {Mo}}, \bibinfo {author} {\bibfnamefont {K.~T.}\ \bibnamefont {Turner}}, \
  and\ \bibinfo {author} {\bibfnamefont {I.}~\bibnamefont {Szlufarska}},\
  }\href@noop {} {\bibfield  {journal} {\bibinfo  {journal} {Nature}\ }\textbf
  {\bibinfo {volume} {457}},\ \bibinfo {pages} {1116} (\bibinfo {year}
  {2009})}\BibitemShut {NoStop}%
\bibitem [{\citenamefont {Landman}\ \emph {et~al.}(1990)\citenamefont
  {Landman}, \citenamefont {Luedtke}, \citenamefont {Burnham},\ and\
  \citenamefont {Colton}}]{land90}%
  \BibitemOpen
  \bibfield  {author} {\bibinfo {author} {\bibfnamefont {U.}~\bibnamefont
  {Landman}}, \bibinfo {author} {\bibfnamefont {W.}~\bibnamefont {Luedtke}},
  \bibinfo {author} {\bibfnamefont {N.}~\bibnamefont {Burnham}}, \ and\
  \bibinfo {author} {\bibfnamefont {R.~J.}\ \bibnamefont {Colton}},\
  }\href@noop {} {\bibfield  {journal} {\bibinfo  {journal} {Science}\ }\textbf
  {\bibinfo {volume} {248}},\ \bibinfo {pages} {454} (\bibinfo {year}
  {1990})}\BibitemShut {NoStop}%
\bibitem [{\citenamefont {Brant}\ and\ \citenamefont
  {Childress}(2004)}]{bran04}%
  \BibitemOpen
  \bibfield  {author} {\bibinfo {author} {\bibfnamefont {J.~A.}\ \bibnamefont
  {Brant}}\ and\ \bibinfo {author} {\bibfnamefont {A.~E.}\ \bibnamefont
  {Childress}},\ }\href@noop {} {\bibfield  {journal} {\bibinfo  {journal} {J
  Membrane Sci.}\ }\textbf {\bibinfo {volume} {241}},\ \bibinfo {pages} {235}
  (\bibinfo {year} {2004})}\BibitemShut {NoStop}%
\bibitem [{\citenamefont {Johnson}\ \emph {et~al.}(1971)\citenamefont
  {Johnson}, \citenamefont {Kendall},\ and\ \citenamefont {Roberts}}]{john71}%
  \BibitemOpen
  \bibfield  {author} {\bibinfo {author} {\bibfnamefont {K.}~\bibnamefont
  {Johnson}}, \bibinfo {author} {\bibfnamefont {K.}~\bibnamefont {Kendall}}, \
  and\ \bibinfo {author} {\bibfnamefont {A.}~\bibnamefont {Roberts}},\
  }\href@noop {} {\bibfield  {journal} {\bibinfo  {journal} {Proc. Roy. Soc.
  A}\ }\textbf {\bibinfo {volume} {324}},\ \bibinfo {pages} {301} (\bibinfo
  {year} {1971})}\BibitemShut {NoStop}%
\bibitem [{\citenamefont {Mary}\ \emph {et~al.}(2006)\citenamefont {Mary},
  \citenamefont {Chateauminois},\ and\ \citenamefont {Fretigny}}]{mary06}%
  \BibitemOpen
  \bibfield  {author} {\bibinfo {author} {\bibfnamefont {P.}~\bibnamefont
  {Mary}}, \bibinfo {author} {\bibfnamefont {A.}~\bibnamefont {Chateauminois}},
  \ and\ \bibinfo {author} {\bibfnamefont {C.}~\bibnamefont {Fretigny}},\
  }\href@noop {} {\bibfield  {journal} {\bibinfo  {journal} {J. Phys. D: Appl.
  Phys.}\ }\textbf {\bibinfo {volume} {39}},\ \bibinfo {pages} {3665} (\bibinfo
  {year} {2006})}\BibitemShut {NoStop}%
\bibitem [{\citenamefont {Chateauminois}\ \emph {et~al.}(2010)\citenamefont
  {Chateauminois}, \citenamefont {Fretigny},\ and\ \citenamefont
  {Olanier}}]{chat10}%
  \BibitemOpen
  \bibfield  {author} {\bibinfo {author} {\bibfnamefont {A.}~\bibnamefont
  {Chateauminois}}, \bibinfo {author} {\bibfnamefont {C.}~\bibnamefont
  {Fretigny}}, \ and\ \bibinfo {author} {\bibfnamefont {L.}~\bibnamefont
  {Olanier}},\ }\href@noop {} {\bibfield  {journal} {\bibinfo  {journal} {Phys.
  Rev. E}\ }\textbf {\bibinfo {volume} {81}},\ \bibinfo {pages} {026106}
  (\bibinfo {year} {2010})}\BibitemShut {NoStop}%
\bibitem [{\citenamefont {Butt}\ \emph {et~al.}(2005)\citenamefont {Butt},
  \citenamefont {Cappella},\ and\ \citenamefont {Kappl}}]{butt05}%
  \BibitemOpen
  \bibfield  {author} {\bibinfo {author} {\bibfnamefont {H.-J.}\ \bibnamefont
  {Butt}}, \bibinfo {author} {\bibfnamefont {B.}~\bibnamefont {Cappella}}, \
  and\ \bibinfo {author} {\bibfnamefont {M.}~\bibnamefont {Kappl}},\
  }\href@noop {} {\bibfield  {journal} {\bibinfo  {journal} {Surf. Sci. Rep.}\
  }\textbf {\bibinfo {volume} {59}},\ \bibinfo {pages} {1} (\bibinfo {year}
  {2005})}\BibitemShut {NoStop}%
\bibitem [{\citenamefont {Erath}\ \emph {et~al.}(2010)\citenamefont {Erath},
  \citenamefont {Schmidt},\ and\ \citenamefont {Fery}}]{erat10}%
  \BibitemOpen
  \bibfield  {author} {\bibinfo {author} {\bibfnamefont {J.}~\bibnamefont
  {Erath}}, \bibinfo {author} {\bibfnamefont {S.}~\bibnamefont {Schmidt}}, \
  and\ \bibinfo {author} {\bibfnamefont {A.}~\bibnamefont {Fery}},\ }\href@noop
  {} {\bibfield  {journal} {\bibinfo  {journal} {Soft Matter}\ }\textbf
  {\bibinfo {volume} {6}},\ \bibinfo {pages} {1432} (\bibinfo {year}
  {2010})}\BibitemShut {NoStop}%
\bibitem [{\citenamefont {Noy}\ \emph {et~al.}(1997)\citenamefont {Noy},
  \citenamefont {Vezenov},\ and\ \citenamefont {Lieber}}]{noy97}%
  \BibitemOpen
  \bibfield  {author} {\bibinfo {author} {\bibfnamefont {A.}~\bibnamefont
  {Noy}}, \bibinfo {author} {\bibfnamefont {D.~V.}\ \bibnamefont {Vezenov}}, \
  and\ \bibinfo {author} {\bibfnamefont {C.~M.}\ \bibnamefont {Lieber}},\
  }\href@noop {} {\bibfield  {journal} {\bibinfo  {journal} {Ann Rev. Mater.
  Sci.}\ }\textbf {\bibinfo {volume} {27}},\ \bibinfo {pages} {381} (\bibinfo
  {year} {1997})}\BibitemShut {NoStop}%
\bibitem [{\citenamefont {de~Gennes}\ \emph {et~al.}(2004)\citenamefont
  {de~Gennes}, \citenamefont {Brochard-Wyart},\ and\ \citenamefont
  {Quere}}]{dege04}%
  \BibitemOpen
  \bibfield  {author} {\bibinfo {author} {\bibfnamefont {P.-G.}\ \bibnamefont
  {de~Gennes}}, \bibinfo {author} {\bibfnamefont {F.}~\bibnamefont
  {Brochard-Wyart}}, \ and\ \bibinfo {author} {\bibfnamefont {D.}~\bibnamefont
  {Quere}},\ }\href@noop {} {\emph {\bibinfo {title} {Capillarity and Wetting
  Phenomena: Drops, Bubbles, Pearls, Waves}}}\ (\bibinfo  {publisher}
  {Springer},\ \bibinfo {year} {2004})\BibitemShut {NoStop}%
\bibitem [{\citenamefont {Bhushan}\ \emph {et~al.}(1995)\citenamefont
  {Bhushan}, \citenamefont {Israelachvili},\ and\ \citenamefont
  {Landman}}]{bhus95}%
  \BibitemOpen
  \bibfield  {author} {\bibinfo {author} {\bibfnamefont {B.}~\bibnamefont
  {Bhushan}}, \bibinfo {author} {\bibfnamefont {J.~N.}\ \bibnamefont
  {Israelachvili}}, \ and\ \bibinfo {author} {\bibfnamefont {U.}~\bibnamefont
  {Landman}},\ }\href@noop {} {\bibfield  {journal} {\bibinfo  {journal}
  {Nature}\ }\textbf {\bibinfo {volume} {374}},\ \bibinfo {pages} {607}
  (\bibinfo {year} {1995})}\BibitemShut {NoStop}%
\bibitem [{\citenamefont {DelRio}\ \emph {et~al.}(2005)\citenamefont {DelRio},
  \citenamefont {de~Boer}, \citenamefont {Knapp}, \citenamefont {Reedy},
  \citenamefont {Clews},\ and\ \citenamefont {Dunn}}]{delr05}%
  \BibitemOpen
  \bibfield  {author} {\bibinfo {author} {\bibfnamefont {F.~W.}\ \bibnamefont
  {DelRio}}, \bibinfo {author} {\bibfnamefont {M.~P.}\ \bibnamefont {de~Boer}},
  \bibinfo {author} {\bibfnamefont {J.~A.}\ \bibnamefont {Knapp}}, \bibinfo
  {author} {\bibfnamefont {E.~D.}\ \bibnamefont {Reedy}}, \bibinfo {author}
  {\bibfnamefont {P.~J.}\ \bibnamefont {Clews}}, \ and\ \bibinfo {author}
  {\bibfnamefont {M.~L.}\ \bibnamefont {Dunn}},\ }\href@noop {} {\bibfield
  {journal} {\bibinfo  {journal} {Nature Mater.}\ }\textbf {\bibinfo {volume}
  {4}},\ \bibinfo {pages} {629} (\bibinfo {year} {2005})}\BibitemShut {NoStop}%
\bibitem [{\citenamefont {Long}\ \emph {et~al.}(1996)\citenamefont {Long},
  \citenamefont {Ajdari},\ and\ \citenamefont {Leibler}}]{long96}%
  \BibitemOpen
  \bibfield  {author} {\bibinfo {author} {\bibfnamefont {D.}~\bibnamefont
  {Long}}, \bibinfo {author} {\bibfnamefont {A.}~\bibnamefont {Ajdari}}, \ and\
  \bibinfo {author} {\bibfnamefont {L.}~\bibnamefont {Leibler}},\ }\href@noop
  {} {\bibfield  {journal} {\bibinfo  {journal} {Langmuir}\ }\textbf {\bibinfo
  {volume} {12}},\ \bibinfo {pages} {5221} (\bibinfo {year}
  {1996})}\BibitemShut {NoStop}%
\bibitem [{\citenamefont {Jerison}\ \emph {et~al.}(2011)\citenamefont
  {Jerison}, \citenamefont {Xu}, \citenamefont {Wilen},\ and\ \citenamefont
  {Dufresne}}]{jeri11}%
  \BibitemOpen
  \bibfield  {author} {\bibinfo {author} {\bibfnamefont {E.~R.}\ \bibnamefont
  {Jerison}}, \bibinfo {author} {\bibfnamefont {Y.}~\bibnamefont {Xu}},
  \bibinfo {author} {\bibfnamefont {L.~A.}\ \bibnamefont {Wilen}}, \ and\
  \bibinfo {author} {\bibfnamefont {E.~R.}\ \bibnamefont {Dufresne}},\
  }\href@noop {} {\bibfield  {journal} {\bibinfo  {journal} {Phys. Rev. Lett.}\
  }\textbf {\bibinfo {volume} {106}},\ \bibinfo {pages} {186103} (\bibinfo
  {year} {2011})}\BibitemShut {NoStop}%
\bibitem [{\citenamefont {Style}\ and\ \citenamefont
  {Dufresne}(2012)}]{styl12}%
  \BibitemOpen
  \bibfield  {author} {\bibinfo {author} {\bibfnamefont {R.~W.}\ \bibnamefont
  {Style}}\ and\ \bibinfo {author} {\bibfnamefont {E.~R.}\ \bibnamefont
  {Dufresne}},\ }\href@noop {} {\bibfield  {journal} {\bibinfo  {journal} {Soft
  Matter}\ }\textbf {\bibinfo {volume} {8}},\ \bibinfo {pages} {7177} (\bibinfo
  {year} {2012})}\BibitemShut {NoStop}%
\bibitem [{\citenamefont {Style}\ \emph
  {et~al.}(2013{\natexlab{a}})\citenamefont {Style}, \citenamefont
  {Boltyanskiy}, \citenamefont {Che}, \citenamefont {Wettlaufer}, \citenamefont
  {Wilen},\ and\ \citenamefont {Dufresne}}]{styl12c}%
  \BibitemOpen
  \bibfield  {author} {\bibinfo {author} {\bibfnamefont {R.~W.}\ \bibnamefont
  {Style}}, \bibinfo {author} {\bibfnamefont {R.}~\bibnamefont {Boltyanskiy}},
  \bibinfo {author} {\bibfnamefont {Y.}~\bibnamefont {Che}}, \bibinfo {author}
  {\bibfnamefont {J.~S.}\ \bibnamefont {Wettlaufer}}, \bibinfo {author}
  {\bibfnamefont {L.~A.}\ \bibnamefont {Wilen}}, \ and\ \bibinfo {author}
  {\bibfnamefont {E.~R.}\ \bibnamefont {Dufresne}},\ }\href@noop {} {\bibfield
  {journal} {\bibinfo  {journal} {Phys. Rev. Lett.}\ }\textbf {\bibinfo
  {volume} {110}},\ \bibinfo {pages} {066103} (\bibinfo {year}
  {2013}{\natexlab{a}})}\BibitemShut {NoStop}%
\bibitem [{\citenamefont {Style}\ \emph
  {et~al.}(2013{\natexlab{b}})\citenamefont {Style}, \citenamefont {Che},
  \citenamefont {Park}, \citenamefont {Weon}, \citenamefont {Je}, \citenamefont
  {Hyland}, \citenamefont {German}, \citenamefont {Power}, \citenamefont
  {Wilen}, \citenamefont {Wettlaufer},\ and\ \citenamefont
  {Dufresne}}]{styl13}%
  \BibitemOpen
  \bibfield  {author} {\bibinfo {author} {\bibfnamefont {R.~W.}\ \bibnamefont
  {Style}}, \bibinfo {author} {\bibfnamefont {Y.}~\bibnamefont {Che}}, \bibinfo
  {author} {\bibfnamefont {S.~J.}\ \bibnamefont {Park}}, \bibinfo {author}
  {\bibfnamefont {B.~M.}\ \bibnamefont {Weon}}, \bibinfo {author}
  {\bibfnamefont {J.~H.}\ \bibnamefont {Je}}, \bibinfo {author} {\bibfnamefont
  {C.}~\bibnamefont {Hyland}}, \bibinfo {author} {\bibfnamefont {G.~K.}\
  \bibnamefont {German}}, \bibinfo {author} {\bibfnamefont {M.~P.}\
  \bibnamefont {Power}}, \bibinfo {author} {\bibfnamefont {L.~A.}\ \bibnamefont
  {Wilen}}, \bibinfo {author} {\bibfnamefont {J.~S.}\ \bibnamefont
  {Wettlaufer}}, \ and\ \bibinfo {author} {\bibfnamefont {E.~R.}\ \bibnamefont
  {Dufresne}},\ }\href@noop {} {\bibfield  {journal} {\bibinfo  {journal}
  {Proc. Nat. Acad. Sci.}\ }\textbf {\bibinfo {volume} {110}},\ \bibinfo
  {pages} {12541} (\bibinfo {year} {2013}{\natexlab{b}})}\BibitemShut {NoStop}%
\bibitem [{\citenamefont {Chakrabarti}\ and\ \citenamefont
  {Chaudhury}(2013)}]{chak13}%
  \BibitemOpen
  \bibfield  {author} {\bibinfo {author} {\bibfnamefont {A.}~\bibnamefont
  {Chakrabarti}}\ and\ \bibinfo {author} {\bibfnamefont {M.~K.}\ \bibnamefont
  {Chaudhury}},\ }\href@noop {} {\bibfield  {journal} {\bibinfo  {journal}
  {Langmuir}\ }\textbf {\bibinfo {volume} {0}},\ \bibinfo {pages} {In Press}
  (\bibinfo {year} {2013})}\BibitemShut {NoStop}%
\bibitem [{\citenamefont {Mora}\ \emph {et~al.}(2013)\citenamefont {Mora},
  \citenamefont {Maurini}, \citenamefont {Phou}, \citenamefont {Fromental},
  \citenamefont {Audoly},\ and\ \citenamefont {Pomeau}}]{mora13}%
  \BibitemOpen
  \bibfield  {author} {\bibinfo {author} {\bibfnamefont {S.}~\bibnamefont
  {Mora}}, \bibinfo {author} {\bibfnamefont {C.}~\bibnamefont {Maurini}},
  \bibinfo {author} {\bibfnamefont {T.}~\bibnamefont {Phou}}, \bibinfo {author}
  {\bibfnamefont {J.-M.}\ \bibnamefont {Fromental}}, \bibinfo {author}
  {\bibfnamefont {B.}~\bibnamefont {Audoly}}, \ and\ \bibinfo {author}
  {\bibfnamefont {Y.}~\bibnamefont {Pomeau}},\ }\href@noop {} {\bibfield
  {journal} {\bibinfo  {journal} {Phys. Rev. Lett.}\ }\textbf {\bibinfo
  {volume} {111}},\ \bibinfo {pages} {114301} (\bibinfo {year}
  {2013})}\BibitemShut {NoStop}%
\bibitem [{\citenamefont {Mora}\ \emph {et~al.}(2010)\citenamefont {Mora},
  \citenamefont {Phou}, \citenamefont {Fromental}, \citenamefont {Pismen},\
  and\ \citenamefont {Pomeau}}]{mora10}%
  \BibitemOpen
  \bibfield  {author} {\bibinfo {author} {\bibfnamefont {S.}~\bibnamefont
  {Mora}}, \bibinfo {author} {\bibfnamefont {T.}~\bibnamefont {Phou}}, \bibinfo
  {author} {\bibfnamefont {J.-M.}\ \bibnamefont {Fromental}}, \bibinfo {author}
  {\bibfnamefont {L.~M.}\ \bibnamefont {Pismen}}, \ and\ \bibinfo {author}
  {\bibfnamefont {Y.}~\bibnamefont {Pomeau}},\ }\href@noop {} {\bibfield
  {journal} {\bibinfo  {journal} {Phys. Rev. Lett.}\ }\textbf {\bibinfo
  {volume} {105}},\ \bibinfo {pages} {214301} (\bibinfo {year}
  {2010})}\BibitemShut {NoStop}%
\bibitem [{\citenamefont {Gordan}\ \emph {et~al.}(2008)\citenamefont {Gordan},
  \citenamefont {Persson}, \citenamefont {Cesa}, \citenamefont {Mayer},
  \citenamefont {Hoffmann}, \citenamefont {Dieluweit},\ and\ \citenamefont
  {Merkel}}]{gord08}%
  \BibitemOpen
  \bibfield  {author} {\bibinfo {author} {\bibfnamefont {O.~D.}\ \bibnamefont
  {Gordan}}, \bibinfo {author} {\bibfnamefont {B.~N.}\ \bibnamefont {Persson}},
  \bibinfo {author} {\bibfnamefont {C.~M.}\ \bibnamefont {Cesa}}, \bibinfo
  {author} {\bibfnamefont {D.}~\bibnamefont {Mayer}}, \bibinfo {author}
  {\bibfnamefont {B.}~\bibnamefont {Hoffmann}}, \bibinfo {author}
  {\bibfnamefont {S.}~\bibnamefont {Dieluweit}}, \ and\ \bibinfo {author}
  {\bibfnamefont {R.}~\bibnamefont {Merkel}},\ }\href@noop {} {\bibfield
  {journal} {\bibinfo  {journal} {Langmuir}\ }\textbf {\bibinfo {volume}
  {24}},\ \bibinfo {pages} {6636} (\bibinfo {year} {2008})}\BibitemShut
  {NoStop}%
\bibitem [{\citenamefont {Persson}\ \emph {et~al.}(2010)\citenamefont
  {Persson}, \citenamefont {Kovalev}, \citenamefont {Wasem}, \citenamefont
  {Gnecco},\ and\ \citenamefont {Gorb}}]{pers10}%
  \BibitemOpen
  \bibfield  {author} {\bibinfo {author} {\bibfnamefont {B.}~\bibnamefont
  {Persson}}, \bibinfo {author} {\bibfnamefont {A.}~\bibnamefont {Kovalev}},
  \bibinfo {author} {\bibfnamefont {M.}~\bibnamefont {Wasem}}, \bibinfo
  {author} {\bibfnamefont {E.}~\bibnamefont {Gnecco}}, \ and\ \bibinfo {author}
  {\bibfnamefont {S.}~\bibnamefont {Gorb}},\ }\href@noop {} {\bibfield
  {journal} {\bibinfo  {journal} {Europhys. Lett.}\ }\textbf {\bibinfo {volume}
  {92}},\ \bibinfo {pages} {46001} (\bibinfo {year} {2010})}\BibitemShut
  {NoStop}%
\bibitem [{\citenamefont {Jagota}\ \emph {et~al.}(2012)\citenamefont {Jagota},
  \citenamefont {Paretkar},\ and\ \citenamefont {Ghatak}}]{jago12}%
  \BibitemOpen
  \bibfield  {author} {\bibinfo {author} {\bibfnamefont {A.}~\bibnamefont
  {Jagota}}, \bibinfo {author} {\bibfnamefont {D.}~\bibnamefont {Paretkar}}, \
  and\ \bibinfo {author} {\bibfnamefont {A.}~\bibnamefont {Ghatak}},\
  }\href@noop {} {\bibfield  {journal} {\bibinfo  {journal} {Phys. Rev. E}\
  }\textbf {\bibinfo {volume} {85}},\ \bibinfo {pages} {051602} (\bibinfo
  {year} {2012})}\BibitemShut {NoStop}%
\bibitem [{\citenamefont {Mora}\ \emph {et~al.}(2011)\citenamefont {Mora},
  \citenamefont {Abkarian}, \citenamefont {Tabuteau},\ and\ \citenamefont
  {Pomeau}}]{mora11}%
  \BibitemOpen
  \bibfield  {author} {\bibinfo {author} {\bibfnamefont {S.}~\bibnamefont
  {Mora}}, \bibinfo {author} {\bibfnamefont {M.}~\bibnamefont {Abkarian}},
  \bibinfo {author} {\bibfnamefont {H.}~\bibnamefont {Tabuteau}}, \ and\
  \bibinfo {author} {\bibfnamefont {Y.}~\bibnamefont {Pomeau}},\ }\href@noop {}
  {\bibfield  {journal} {\bibinfo  {journal} {Soft Matter}\ }\textbf {\bibinfo
  {volume} {7}},\ \bibinfo {pages} {10612} (\bibinfo {year}
  {2011})}\BibitemShut {NoStop}%
\bibitem [{\citenamefont {Chen}\ \emph {et~al.}(2012)\citenamefont {Chen},
  \citenamefont {Cai}, \citenamefont {Suo},\ and\ \citenamefont
  {Hayward}}]{chen12}%
  \BibitemOpen
  \bibfield  {author} {\bibinfo {author} {\bibfnamefont {D.}~\bibnamefont
  {Chen}}, \bibinfo {author} {\bibfnamefont {S.}~\bibnamefont {Cai}}, \bibinfo
  {author} {\bibfnamefont {Z.}~\bibnamefont {Suo}}, \ and\ \bibinfo {author}
  {\bibfnamefont {R.~C.}\ \bibnamefont {Hayward}},\ }\href@noop {} {\bibfield
  {journal} {\bibinfo  {journal} {Phys. Rev. Lett.}\ }\textbf {\bibinfo
  {volume} {109}},\ \bibinfo {pages} {038001} (\bibinfo {year}
  {2012})}\BibitemShut {NoStop}%
\bibitem [{\citenamefont {Lee}\ \emph {et~al.}(2012)\citenamefont {Lee},
  \citenamefont {Peng}, \citenamefont {Yang}, \citenamefont {Roh},
  \citenamefont {Funabashi}, \citenamefont {Park}, \citenamefont {Rice},
  \citenamefont {Chen}, \citenamefont {Long}, \citenamefont {Wu} \emph
  {et~al.}}]{lee12}%
  \BibitemOpen
  \bibfield  {author} {\bibinfo {author} {\bibfnamefont {J.~B.}\ \bibnamefont
  {Lee}}, \bibinfo {author} {\bibfnamefont {S.}~\bibnamefont {Peng}}, \bibinfo
  {author} {\bibfnamefont {D.}~\bibnamefont {Yang}}, \bibinfo {author}
  {\bibfnamefont {Y.~H.}\ \bibnamefont {Roh}}, \bibinfo {author} {\bibfnamefont
  {H.}~\bibnamefont {Funabashi}}, \bibinfo {author} {\bibfnamefont
  {N.}~\bibnamefont {Park}}, \bibinfo {author} {\bibfnamefont {E.~J.}\
  \bibnamefont {Rice}}, \bibinfo {author} {\bibfnamefont {L.}~\bibnamefont
  {Chen}}, \bibinfo {author} {\bibfnamefont {R.}~\bibnamefont {Long}}, \bibinfo
  {author} {\bibfnamefont {M.}~\bibnamefont {Wu}},  \emph {et~al.},\
  }\href@noop {} {\bibfield  {journal} {\bibinfo  {journal} {Nature Nano.}\
  }\textbf {\bibinfo {volume} {7}},\ \bibinfo {pages} {816} (\bibinfo {year}
  {2012})}\BibitemShut {NoStop}%
\bibitem [{\citenamefont {Xu}\ \emph {et~al.}(2013)\citenamefont {Xu},
  \citenamefont {Jagota}, \citenamefont {Peng}, \citenamefont {Luo},
  \citenamefont {Wu},\ and\ \citenamefont {Hui}}]{xu13}%
  \BibitemOpen
  \bibfield  {author} {\bibinfo {author} {\bibfnamefont {X.}~\bibnamefont
  {Xu}}, \bibinfo {author} {\bibfnamefont {A.}~\bibnamefont {Jagota}}, \bibinfo
  {author} {\bibfnamefont {S.}~\bibnamefont {Peng}}, \bibinfo {author}
  {\bibfnamefont {D.}~\bibnamefont {Luo}}, \bibinfo {author} {\bibfnamefont
  {M.}~\bibnamefont {Wu}}, \ and\ \bibinfo {author} {\bibfnamefont {C.-Y.}\
  \bibnamefont {Hui}},\ }\href@noop {} {\bibfield  {journal} {\bibinfo
  {journal} {Langmuir}\ }\textbf {\bibinfo {volume} {29}},\ \bibinfo {pages}
  {8665} (\bibinfo {year} {2013})}\BibitemShut {NoStop}%
\bibitem [{\citenamefont {Hertz}(1882)}]{hert82}%
  \BibitemOpen
  \bibfield  {author} {\bibinfo {author} {\bibfnamefont {H.}~\bibnamefont
  {Hertz}},\ }\href@noop {} {\bibfield  {journal} {\bibinfo  {journal} {J.
  Reine Angew. Math.}\ }\textbf {\bibinfo {volume} {92}},\ \bibinfo {pages}
  {156} (\bibinfo {year} {1882})}\BibitemShut {NoStop}%
\bibitem [{\citenamefont {Frenkel}(1945)}]{fren45}%
  \BibitemOpen
  \bibfield  {author} {\bibinfo {author} {\bibfnamefont {J.}~\bibnamefont
  {Frenkel}},\ }\href@noop {} {\bibfield  {journal} {\bibinfo  {journal} {J.
  Phys}\ }\textbf {\bibinfo {volume} {9}},\ \bibinfo {pages} {385} (\bibinfo
  {year} {1945})}\BibitemShut {NoStop}%
\bibitem [{\citenamefont {Jagota}\ \emph {et~al.}(1998)\citenamefont {Jagota},
  \citenamefont {Argento},\ and\ \citenamefont {Mazur}}]{jago98}%
  \BibitemOpen
  \bibfield  {author} {\bibinfo {author} {\bibfnamefont {A.}~\bibnamefont
  {Jagota}}, \bibinfo {author} {\bibfnamefont {C.}~\bibnamefont {Argento}}, \
  and\ \bibinfo {author} {\bibfnamefont {S.}~\bibnamefont {Mazur}},\
  }\href@noop {} {\bibfield  {journal} {\bibinfo  {journal} {J. Appl. Phys.}\
  }\textbf {\bibinfo {volume} {83}},\ \bibinfo {pages} {250} (\bibinfo {year}
  {1998})}\BibitemShut {NoStop}%
\bibitem [{\citenamefont {Lin}\ \emph {et~al.}(2001)\citenamefont {Lin},
  \citenamefont {Hui},\ and\ \citenamefont {Jagota}}]{lin01}%
  \BibitemOpen
  \bibfield  {author} {\bibinfo {author} {\bibfnamefont {Y.}~\bibnamefont
  {Lin}}, \bibinfo {author} {\bibfnamefont {C.}~\bibnamefont {Hui}}, \ and\
  \bibinfo {author} {\bibfnamefont {A.}~\bibnamefont {Jagota}},\ }\href@noop {}
  {\bibfield  {journal} {\bibinfo  {journal} {J. Colloid Interface Sci,}\
  }\textbf {\bibinfo {volume} {237}},\ \bibinfo {pages} {267} (\bibinfo {year}
  {2001})}\BibitemShut {NoStop}%
\bibitem [{\citenamefont {Maugis}(1995)}]{maug95}%
  \BibitemOpen
  \bibfield  {author} {\bibinfo {author} {\bibfnamefont {D.}~\bibnamefont
  {Maugis}},\ }\href@noop {} {\bibfield  {journal} {\bibinfo  {journal}
  {Langmuir}\ }\textbf {\bibinfo {volume} {11}},\ \bibinfo {pages} {679}
  (\bibinfo {year} {1995})}\BibitemShut {NoStop}%
\bibitem [{\citenamefont {Tabor}(1977)}]{tabo77}%
  \BibitemOpen
  \bibfield  {author} {\bibinfo {author} {\bibfnamefont {D.}~\bibnamefont
  {Tabor}},\ }\href@noop {} {\bibfield  {journal} {\bibinfo  {journal} {J.
  Colloid Interface Sci.}\ }\textbf {\bibinfo {volume} {58}},\ \bibinfo {pages}
  {2} (\bibinfo {year} {1977})}\BibitemShut {NoStop}%
\bibitem [{\citenamefont {Maugis}(1992)}]{maug92}%
  \BibitemOpen
  \bibfield  {author} {\bibinfo {author} {\bibfnamefont {D.}~\bibnamefont
  {Maugis}},\ }\href@noop {} {\bibfield  {journal} {\bibinfo  {journal} {J.
  Colloid Interface Sci.}\ }\textbf {\bibinfo {volume} {150}},\ \bibinfo
  {pages} {243} (\bibinfo {year} {1992})}\BibitemShut {NoStop}%
\bibitem [{\citenamefont {Carrillo}\ \emph {et~al.}(2010)\citenamefont
  {Carrillo}, \citenamefont {Raphael},\ and\ \citenamefont
  {Dobrynin}}]{carr10}%
  \BibitemOpen
  \bibfield  {author} {\bibinfo {author} {\bibfnamefont {J.-M.~Y.}\
  \bibnamefont {Carrillo}}, \bibinfo {author} {\bibfnamefont {E.}~\bibnamefont
  {Raphael}}, \ and\ \bibinfo {author} {\bibfnamefont {A.~V.}\ \bibnamefont
  {Dobrynin}},\ }\href@noop {} {\bibfield  {journal} {\bibinfo  {journal}
  {Langmuir}\ }\textbf {\bibinfo {volume} {26}},\ \bibinfo {pages} {12973}
  (\bibinfo {year} {2010})}\BibitemShut {NoStop}%
\bibitem [{\citenamefont {Shuttleworth}(1950)}]{shut50}%
  \BibitemOpen
  \bibfield  {author} {\bibinfo {author} {\bibfnamefont {R.}~\bibnamefont
  {Shuttleworth}},\ }\href@noop {} {\bibfield  {journal} {\bibinfo  {journal}
  {Proc. Phys. Soc. A}\ }\textbf {\bibinfo {volume} {63}},\ \bibinfo {pages}
  {444} (\bibinfo {year} {1950})}\BibitemShut {NoStop}%
\bibitem [{\citenamefont {Hui}\ and\ \citenamefont {Jagota}(2013)}]{hui13}%
  \BibitemOpen
  \bibfield  {author} {\bibinfo {author} {\bibfnamefont {C.-Y.}\ \bibnamefont
  {Hui}}\ and\ \bibinfo {author} {\bibfnamefont {A.}~\bibnamefont {Jagota}},\
  }\href@noop {} {\bibfield  {journal} {\bibinfo  {journal} {Langmuir}\
  }\textbf {\bibinfo {volume} {29}},\ \bibinfo {pages} {11310} (\bibinfo {year}
  {2013})}\BibitemShut {NoStop}%
\bibitem [{\citenamefont {Nadermann}\ \emph {et~al.}(2013)\citenamefont
  {Nadermann}, \citenamefont {Hui},\ and\ \citenamefont {Jagota}}]{nade13}%
  \BibitemOpen
  \bibfield  {author} {\bibinfo {author} {\bibfnamefont {N.}~\bibnamefont
  {Nadermann}}, \bibinfo {author} {\bibfnamefont {C.-Y.}\ \bibnamefont {Hui}},
  \ and\ \bibinfo {author} {\bibfnamefont {A.}~\bibnamefont {Jagota}},\
  }\href@noop {} {\bibfield  {journal} {\bibinfo  {journal} {Proc. Nat. Acad.
  Sci.}\ }\textbf {\bibinfo {volume} {110}},\ \bibinfo {pages} {10541}
  (\bibinfo {year} {2013})}\BibitemShut {NoStop}%
\bibitem [{\citenamefont {Schimmele}\ \emph {et~al.}(2007)\citenamefont
  {Schimmele}, \citenamefont {Napi{\'o}rkowski},\ and\ \citenamefont
  {Dietrich}}]{schi07}%
  \BibitemOpen
  \bibfield  {author} {\bibinfo {author} {\bibfnamefont {L.}~\bibnamefont
  {Schimmele}}, \bibinfo {author} {\bibfnamefont {M.}~\bibnamefont
  {Napi{\'o}rkowski}}, \ and\ \bibinfo {author} {\bibfnamefont
  {S.}~\bibnamefont {Dietrich}},\ }\href@noop {} {\bibfield  {journal}
  {\bibinfo  {journal} {J. Chem. Phys.}\ }\textbf {\bibinfo {volume} {127}},\
  \bibinfo {pages} {164715} (\bibinfo {year} {2007})}\BibitemShut {NoStop}%
\bibitem [{\citenamefont {Chen}\ \emph {et~al.}(2013)\citenamefont {Chen},
  \citenamefont {Bonaccurso},\ and\ \citenamefont {Shanahan}}]{chen13}%
  \BibitemOpen
  \bibfield  {author} {\bibinfo {author} {\bibfnamefont {L.}~\bibnamefont
  {Chen}}, \bibinfo {author} {\bibfnamefont {E.}~\bibnamefont {Bonaccurso}}, \
  and\ \bibinfo {author} {\bibfnamefont {M.~E.}\ \bibnamefont {Shanahan}},\
  }\href@noop {} {\bibfield  {journal} {\bibinfo  {journal} {Langmuir}\
  }\textbf {\bibinfo {volume} {29}},\ \bibinfo {pages} {1893} (\bibinfo {year}
  {2013})}\BibitemShut {NoStop}%
\bibitem [{\citenamefont {Cesa}\ \emph {et~al.}(2007)\citenamefont {Cesa},
  \citenamefont {Kirchgessner}, \citenamefont {Mayer}, \citenamefont {Schwarz},
  \citenamefont {Hoffmann},\ and\ \citenamefont {Merkel}}]{cesa07}%
  \BibitemOpen
  \bibfield  {author} {\bibinfo {author} {\bibfnamefont {C.~M.}\ \bibnamefont
  {Cesa}}, \bibinfo {author} {\bibfnamefont {N.}~\bibnamefont {Kirchgessner}},
  \bibinfo {author} {\bibfnamefont {D.}~\bibnamefont {Mayer}}, \bibinfo
  {author} {\bibfnamefont {U.~S.}\ \bibnamefont {Schwarz}}, \bibinfo {author}
  {\bibfnamefont {B.}~\bibnamefont {Hoffmann}}, \ and\ \bibinfo {author}
  {\bibfnamefont {R.}~\bibnamefont {Merkel}},\ }\href@noop {} {\bibfield
  {journal} {\bibinfo  {journal} {Rev. Sci. Instrum.}\ }\textbf {\bibinfo
  {volume} {78}},\ \bibinfo {pages} {034301} (\bibinfo {year}
  {2007})}\BibitemShut {NoStop}%
\bibitem [{\citenamefont {Fuard}\ \emph {et~al.}(2008)\citenamefont {Fuard},
  \citenamefont {Tzvetkova-Chevolleau}, \citenamefont {Decossas}, \citenamefont
  {Tracqui},\ and\ \citenamefont {Schiavone}}]{fuar08}%
  \BibitemOpen
  \bibfield  {author} {\bibinfo {author} {\bibfnamefont {D.}~\bibnamefont
  {Fuard}}, \bibinfo {author} {\bibfnamefont {T.}~\bibnamefont
  {Tzvetkova-Chevolleau}}, \bibinfo {author} {\bibfnamefont {S.}~\bibnamefont
  {Decossas}}, \bibinfo {author} {\bibfnamefont {P.}~\bibnamefont {Tracqui}}, \
  and\ \bibinfo {author} {\bibfnamefont {P.}~\bibnamefont {Schiavone}},\
  }\href@noop {} {\bibfield  {journal} {\bibinfo  {journal} {Microelectron.
  Eng.}\ }\textbf {\bibinfo {volume} {85}},\ \bibinfo {pages} {1289} (\bibinfo
  {year} {2008})}\BibitemShut {NoStop}%
\bibitem [{\citenamefont {Merkel}\ \emph {et~al.}(2007)\citenamefont {Merkel},
  \citenamefont {Kirchge{\ss}ner}, \citenamefont {Cesa},\ and\ \citenamefont
  {Hoffmann}}]{merk07}%
  \BibitemOpen
  \bibfield  {author} {\bibinfo {author} {\bibfnamefont {R.}~\bibnamefont
  {Merkel}}, \bibinfo {author} {\bibfnamefont {N.}~\bibnamefont
  {Kirchge{\ss}ner}}, \bibinfo {author} {\bibfnamefont {C.~M.}\ \bibnamefont
  {Cesa}}, \ and\ \bibinfo {author} {\bibfnamefont {B.}~\bibnamefont
  {Hoffmann}},\ }\href@noop {} {\bibfield  {journal} {\bibinfo  {journal}
  {Biophys. J.}\ }\textbf {\bibinfo {volume} {93}},\ \bibinfo {pages} {3314}
  (\bibinfo {year} {2007})}\BibitemShut {NoStop}%
\bibitem [{\citenamefont {Xu}\ \emph {et~al.}(2010)\citenamefont {Xu},
  \citenamefont {Engl}, \citenamefont {Jerison}, \citenamefont {Wallenstein},
  \citenamefont {Hyland}, \citenamefont {Wilen},\ and\ \citenamefont
  {Dufresne}}]{xu10}%
  \BibitemOpen
  \bibfield  {author} {\bibinfo {author} {\bibfnamefont {Y.}~\bibnamefont
  {Xu}}, \bibinfo {author} {\bibfnamefont {W.~C.}\ \bibnamefont {Engl}},
  \bibinfo {author} {\bibfnamefont {E.~R.}\ \bibnamefont {Jerison}}, \bibinfo
  {author} {\bibfnamefont {K.~J.}\ \bibnamefont {Wallenstein}}, \bibinfo
  {author} {\bibfnamefont {C.}~\bibnamefont {Hyland}}, \bibinfo {author}
  {\bibfnamefont {L.~A.}\ \bibnamefont {Wilen}}, \ and\ \bibinfo {author}
  {\bibfnamefont {E.~R.}\ \bibnamefont {Dufresne}},\ }\href@noop {} {\bibfield
  {journal} {\bibinfo  {journal} {Proc. Nat. Acad. Sci.}\ }\textbf {\bibinfo
  {volume} {107}},\ \bibinfo {pages} {14964} (\bibinfo {year}
  {2010})}\BibitemShut {NoStop}%
\end{thebibliography}
\end{document}